\documentclass[sigconf,numbers,anonymous=false]{acmart}
\acmConference[ASE]{}{2024}{Sacramento, California, SA}
\usepackage{multirow}    					%multirow
\usepackage{hhline}     				 		%multirow
\usepackage{algorithm} 
\usepackage{algpseudocode}
\usepackage{enumerate}
\usepackage{wrapfig}
\usepackage{fancybox}
\usepackage{url}
\usepackage{balance}
\usepackage{pifont}
\usepackage{xcolor}
\usepackage{tikz}
\usepackage{graphicx}
\usepackage{color}    % \textcolor
\usepackage{listings}
\usepackage{siunitx}
\usepackage{soul} %% for hl
\usepackage[inline]{enumitem} % bring back {description} env

\definecolor{KWColor}{rgb}{0.37,0.08,0.25}
\definecolor{CommentColor}{rgb}{0.133,0.545,0.133}
\definecolor{StringColor}{rgb}{0,0.126,0.941}
\usepackage[utf8]{inputenc}
\usepackage{fancybox}
\usepackage{calc}
\newtheorem{test}{\textbf{Answers to RQ}}

\usepackage{listings}
\usepackage{upquote}
\usepackage{subfigure}
\usepackage{enumitem}
\setlist{leftmargin=2mm}
\usepackage{tcolorbox}
\usepackage{natbib}
\usepackage{fancyvrb,cprotect}

\makeatletter
\def\@fnsymbol#1{\ensuremath{\ifcase#1\or \dagger\or \ddagger\or
   \mathsection\or \mathparagraph\or \|\or **\or \dagger\dagger
   \or \ddagger\ddagger \else\@ctrerr\fi}}
\makeatother

\newcommand{\find}[1]{
\begin{tcolorbox}[leftrule=1mm,toprule=0mm,bottomrule=0mm,left=1pt,right=2pt,top=2pt,bottom=2pt
]
\em #1
\end{tcolorbox}
}

\usepackage{xcolor}
\usepackage{soul}
\soulregister{\cite}7
\soulregister{\citep}7
\soulregister{\citet}7
\soulregister{\ref}7 
\soulregister{\pageref}7
\soulregister{\tool}{7}
\soulregister{\item}7
\soulregister{\subsection}7

\newcommand*\circled[1]{\tikz[baseline=(char.base)]{
            \node[shape=circle,fill,inner sep=1pt] (char) {\textcolor{white}{#1}};}}

\usepackage{algorithm}
\usepackage{algpseudocode}
\usepackage{longtable}

\usepackage{CJK}

\lstdefinestyle{JAVA}{
  language=JAVA,
  moredelim=[is][\underbar]{_}{_},
}

\usepackage{listings}
\usepackage{color}

\definecolor{dkgreen}{rgb}{0,0.6,0}
\definecolor{gray}{rgb}{0.5,0.5,0.5}
\definecolor{mauve}{rgb}{0.58,0,0.82}

\lstdefinestyle{mystyle}{
frame=tb,
  language=Java,
  aboveskip=3mm,
  belowskip=3mm,
  showstringspaces=false,
  columns=flexible,
  basicstyle={\small\ttfamily},
  numbers=none,
  numberstyle=\tiny\color{gray},
  keywordstyle=\color{blue},
  commentstyle=\color{dkgreen},
  stringstyle=\color{mauve},
  breaklines=true,
  breakatwhitespace=true,
  tabsize=3
  escapechar=@
}

\newcommand{\tool}{\textsc{IntelliCiD}}

\usepackage{tikz}
\usetikzlibrary{tikzmark}

\begin{document}

\title{
% Retrospect the Incompatible APIs for Compatibility Issues in Android Applications: An LLM-based Automated Solution
% \tool{}: Towards Automatically Detecting Compatibility Issues in Android Apps
A Large-scale Investigation of Semantically Incompatible APIs behind Compatibility Issues in Android Apps
}

% Authors: Shidong Pan, Tianchen Guo, Lihong Zhang, Xiaoyu Sun*, Zhenchang Xing

\author{Shidong Pan}
\email{shidong.pan@anu.edu.au}
\affiliation{
  \institution{Australian National University \& CSIRO's Data61}
  \country{Australia}
}

\author{Tianchen Guo}
\email{u7439173@anu.edu.au}
\affiliation{
  \institution{Australian National University}
  \country{Australia}
}

\author{Lihong Zhang}
\email{u7261528@anu.edu.au}
\affiliation{
  \institution{Australian National University}
  \country{Australia}
}

\author{Pei Liu}
\email{pei.liu@data61.csiro.au}
\affiliation{
  \institution{CSIRO's Data61}
  \country{Australia}
}

\author{Zhenchang Xing}
\email{zhenchang.xing@data61.csiro.au}
\affiliation{
  \institution{CSIRO's Data61}
  \country{Australia}
}

\author{Xiaoyu Sun}
\authornote{Xiaoyu Sun is the corresponding author.}
\email{xiaoyu.sun1@anu.edu.au}
\affiliation{
  \institution{Australian National University}
  \country{Australia}
}

%------------------------------------------
% use this section to remove copyright info
\setcopyright{none}
\settopmatter{printacmref=false} % Removes citation information below abstract
\renewcommand\footnotetextcopyrightpermission[1]{} % removes footnote with conference information in first column
%------------------------------------------

% adding page numbers
\pagestyle{plain}

\begin{abstract}
Application Programming Interface (API) incompatibility is a long-standing issue in Android application development.
The rapid evolution of Android APIs results in a significant number of API additions, removals, and changes between adjacent versions. 
Unfortunately, this high frequency of alterations may lead to compatibility issues, often without adequate notification to developers regarding these changes. Although researchers have proposed some work on detecting compatibility issues caused by changes in API signatures, they often overlook compatibility issues stemming from sophisticated semantic changes.
In response to this challenge, we conducted a large-scale discovery of incompatible APIs in the Android Open Source Project (AOSP) by leveraging static analysis and pre-trained Large Language Models (LLMs) across adjacent versions. 
We systematically formulate the problem and propose a unified framework to detect incompatible APIs, especially for semantic changes. It's worth highlighting that our approach achieves a 0.83 F1-score in identifying semantically incompatible APIs in the Android framework.
Ultimately, our approach detects 5,481 incompatible APIs spanning from version 4 to version 33. We further demonstrate its effectiveness in supplementing the state-of-the-art methods in detecting a broader spectrum of compatibility issues (+92.3\%) that have been previously overlooked.
\end{abstract}

% \keywords{Compatibility Issues, Android Mobile Apps, Static Analysis, Large Language Models}

\maketitle

\section{Introduction}~\label{sec_intro}
Fragmentation in the Android ecosystem has persistently posed a significant challenge, giving rise to compatibility issues that can result in app crashes on users' Android devices and subsequently degrade the overall user experience~\cite{sun2023taming, li2018cid}. This challenge stems from the fast evolution of Android operation system, which offers thousands of public APIs for developers to access
splendid functionalities, ranging from basic runtime services (e.g., process management, memory management, and device drivers) to hardware facilities~\cite{android_framework_architecture}. Specifically, Google regularly removes or adds APIs to introduce new functionalities or fix bugs. This rapid evolution process may cause potential compatibility issues, leading to abnormal execution or even crashes on Android devices, as indicated by Li et al.~\cite{li2018cid}.

To address this problem, Android formulates a SDK management mechanism that allows app developers to designate the API level on which their apps should run~\cite{wang2017explorative}. With the adoption of SDK management attributes (i.e., \emph{minSdkVersion} and \emph{maxSdkVersion}), developers can specify the certain API levels to avoid incompatible errors on the fly. However, this is insufficient to address all compatibility issues due to the fact that app developers lack comprehensive knowledge about which APIs to safeguard and the corresponding appropriate SDK versions on which the apps can be executed without exception/errors. As evidenced by the various discussions on StackOverflow~\cite{stackoverflownosuchmethoderror, stackoverflowclassnotfound}, even with the SDK management mechanism in place, reports of compatibility issues persistently arise during app execution. 

To identify incompatible Android APIs across different SDK versions, our research fellows have dedicated years of effort to this endeavor, as recently demonstrated in~\cite{li2018cid,wei2016taming,mutchler2016target,zhang2015compatibility,ham2011mobile,huang2018understanding,sun2023taming, sun2022mining}. Indeed, researchers have proposed various static program analysis approaches for characterizing API-induced compatibility issues as APIs are recognized as the set
of  ``entry point''  to the Android ecosystem. For example, Li et al.~\cite{li2018cid} have designed and developed the CiD approach that scans the source code of Android Open Source Project (AOSP)~\cite{sourcecode_aosp} to identify discrepancies on API removals or additions. Wei et al.~\cite{wei2019pivot} further conducted experiments revealing the inconsistency of APIs that are customized by android manufacturers. However, as argued by Sun et al.~\cite{sun2022mining}, most of these static analysis tools only looking into the syntactic changes Android APIs based purely on their signatures, leaving other behavioral changes-induced compatibility issues (a.k.a, semantic compatibility issues) indiscoverable. The behavioral changes have been overlooked mainly because the API implementation changed are way too sophisticated to be handled in a static way~\cite{zhang2022has}. To this end, numerous semantic compatibility issues persist within the Android ecosystem, which are still often exposed at run-time, leading to abnormal output, unexpected exceptions, and even crashes.

Apart from static approaches, there are relatively a few number of dynamic testing tools for tackling compatibility issues on the fly. For example, Sun et al.~\cite{sun2022mining} proposed a test case generation tool named JUnitTestGen, which aims at identifying incompatible APIs at the time
when they are introduced to the framework. In addition, they adopted crowdsourced
testing approach to automatically distribute and execute test cases on real-world
devices to trigger compatibility issues dynamically~\cite{sun2023taming}. Even though some semantic compatibility issues can be triggered dynamically, the capability of dynamic approaches are limited by the poor coverage of Android framework APIs (e.g., the APIs that engaged with UI objects can hardly be constructed programmatically). In addition, dynamic approaches are not guaranteed to trigger all possible compatibility issues with randomly generated inputs. Therefore, detecting compatibility issues in a sound and complete way is a non-trivial task to our community.

To bridge this research gap, we first systematically formulate the incompatible API detection task, then propose a unified framework that attempts to explore compatibility issues in Android system by comprehending the syntax and semantics of API evolution with the assistance of static analysis and pre-trained Large Language Models (LLMs). 
Initially, we formally define the incompatible API, the root cause of compatibility issues, from the perspective of code behaviour evolution instead of API lifespan.
Then, we extract API information from the framework of AOSP\footnote{\url{https://android.googlesource.com/platform/frameworks/base.git}}
%~\pei{Check if the experimental code is this one.} 
spanning API levels from 4 to 33. This initial step is crucial for constructing code facts (such as API signatures, API implementations, etc.) from the Android framework as they serve as the foundation for subsequent compatibility issue analysis.
We further utilize the extracted information to detect incompatible APIs caused by signature and semantic change.
Specifically, we subtly employ LLMs for detecting semantic incompatible APIs, using in-context learning and chain-of-thought strategies.
Our approach achieves an F1-score of 0.83 in on the manually crafted benchmark dataset.
Moreover, we conducted a large-scale investigation of incompatible APIs across all versions from 4 to 33 based on our detection approach.
Among a total of 10,675 sets of APIs that have changed, we identify 5,481 instances of API modifications as potential contributors to compatibility issues.
Armed with the list of incompatible APIs, we supplemented the state-of-the-art method, CiD~\cite{li2018cid}, in detecting a broader spectrum of compatibility issues that had been previously overlooked. Compared to CiD, our approach can detect 92.3\% more compatibility issues induced by semantic API changes.
Additionally, the newly detected semantic APIs are further confirmed by online discussions from Stack Overflow and GitHub.
Overall, we make the following main contributions in this work:
\begin{itemize}
\item We formulate the Android compatibility issues problem as the detection of incompatible APIs.

\item We design and implement a unified framework that leverages the power of the LLM approach to identify incompatible Android framework APIs.

\item We demonstrate the effectiveness of our discovered incompatible APIs in assisting state-of-the-art tools on real-world apps in detecting a wider range of compatibility issues.
\end{itemize}

\section{MOTIVATION}~\label{sec_motivation}
% \xiaoyu{@shidong First, present a code example, and then delve into the difficult nature of static and dynamic tools in identifying semantic continuous integration (CI). And in this scene, we can integrate some introduction of the background about CI. }
Android uses an API level system to ensure mobile apps work across different devices with various operating system versions. As the Android operating system evolves, a new API level is assigned to distinguish specific features and functions newly introduced, accompanying with potential compatibility issues. 
Developers aims to mitigate compatibility issues by actions such as employing version checking in manifest file or forcing version checking (e.g., \texttt{minSdkVersion}, \texttt{targetSdkVersion}) before invoking certain APIs. However, to effectively implement compatibility issues prevention strategies, they need to be aware of which Android APIs are potential incompatible.

\subsection{Empirical Observation}
Compatibility issues have been considered one of the most severe problems in the Android ecosystem. They not only increase the difficulties of developing apps, but also negatively impact the users’ experience, as apps with compatibility issues may not be able to install on users’ devices or may crash at runtime~\cite{liu2021identifying}.
Most of such compatibility issues are caused by incompatible APIs.
Unfortunately, existing methods, such as CiD~\cite{li2018cid}, can only detect incompatible APIs relate to their signatures (i.e., API addition and API removal), but unable to diagnose incompatible APIs relate to syntax and semantics of API evolution.
Listing~\ref{lst_CISemantic} shows an example of API \texttt{getDeviceIds()}.
In Version 15, the API method retrieves the IDs of input devices by interfacing with the window manager (\texttt{IWindowManager}). 
%It does this by first getting an instance of IWindowManager through Display.getWindowManager() and then calling getInputDeviceIds() on this instance. 
If it encounters a RemoteException, it throws a \texttt{RuntimeException} indicating a failure in obtaining input device IDs from the Window Manager.
In Version 16, the implementation is streamlined: the API method directly calls \texttt{getInstance().getInputDeviceIds()}. 
This reflects a design change where the input device IDs are now retrieved directly from the \texttt{InputManager} rather than through the window manager.
Although the change may simplify the process and possibly improving efficiency or reliability, it introduces potential CIs due to its different behaviours between two continuous versions. 
\textbf{Such semantic incompatible APIs cannot be systematically detected by existing tools}, therefore, we propose a novel a unified incompatible API detection framework, especially to tackle this unsolved challenge.
For the aforementioned example, our method can successfully detect out its incompatibility, and correctly report the specific reasons: \textit{Different Return Values} and \textit{Exception Handling Modification}.

\begin{lstlisting}[caption = {A code example of API getDeviceIds(). The API has the semantic change as the exception handling is deleted, which makes it an incompatible API, causing potential compatbility issues.}, label={lst_CISemantic}]
// Version 15, <android.view.InputDevice: int[] getDeviceIds()>
{
    IWindowManager wm = Display.getWindowManager();
    try {
        return wm.getInputDeviceIds();
    } catch (RemoteException ex) {
        throw new RuntimeException("Could not get input device ids from Window Manager.", ex);
    }
}
// Version 16, <android.view.InputDevice: int[] getDeviceIds()>
{
    return InputManager.getInstance().getInputDeviceIds();
}
\end{lstlisting}

%Google has said it will never publish the source code of Android 3.0 (aside from Linux), even though executables have been released to the public. Android 3.1 source code is also being withheld. Thus, Android 3, apart from Linux, is non-free software, pure and simple

\begin{figure}[t]
  \centering  
  \includegraphics[width=\linewidth]{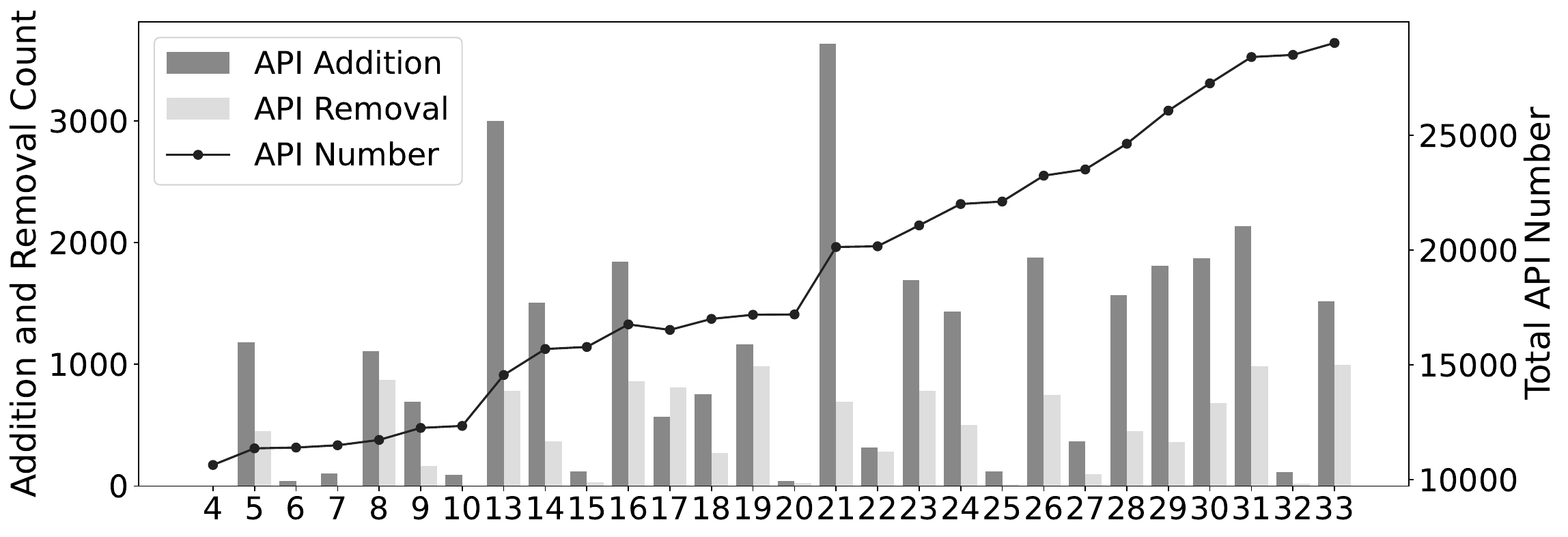}
  \caption{Status Quo of Android public APIs compared to previous version from level 4 to 33. The addition and removal are calculated based on the previous level.}
  \label{fig_APINumbers}
\end{figure}

From 2009 to 2022, Android SDK has updated 29 times from level 4 to level 33\footnote{In this paper, SDK version 11 to 13 are omitted, because Android 3 is non-free software and Google does not provide public available Android 3 source code. According to~\cite{Google_version}}. 
Although the market share of SDK version 4 to 18 is almost negligible (In Android Studio, if you set the \texttt{minSDKVersion} to 19, it would get almost 100\% of devices), we deliberately include them to have a larger sample for the incompatible API exploration and evaluations.
During this period, hundreds of APIs are introduced and deprecated, resulting in a substantial increase in the total number of public APIs – from approximately 10,000 to nearly 30,000, as illustrated in Figure~\ref{fig_APINumbers}. 
% Concurrently, the size of API-related fields expanded approximately sixfold, as detailed in Figure~\ref{fig_FieldNumbers}.
This tremendous growth in the volume of API methods within the Android SDK underscores the critical need for a systematic approach to detect incompatible APIs in the Android framework, highlighting the urgency and importance of this study.

% Manually scrutinizing each API for potential compatibility issues is not only error-prone but also labor-intensive. This process demands a significant investment of time and resources, which is often impractical given the rapid pace of updates and changes in the Android framework.

%-----------------
%-----------------

% %-----------------
% \begin{figure}[t]
%   \centering  
%   \includegraphics[width=\linewidth]{figures/Field_change_bar_line.pdf}
%   \caption{Status Quo of Android Fields compared to previous version from level 4 to 33. The addition and removal are calculated based on the previous level.}
%   \label{fig_FieldNumbers}
% \end{figure}
% %-----------------

\subsection{Problem Statement}~\label{sec_motivation_problem}

Previous research has focused on identifying and detecting CI-related APIs based on their lifespans. 
However, we propose a shift in focus towards the versions in which changes occur, for several reasons:
(1) From a developer's perspective, understanding these version changes can help the implementation of version checking (e.g., \texttt{android.os.Build.VERSION.SDK\_INT >= 23}) or similar CI prevention mechanisms.
(2) It simplifies the process of updating documentation, comments, or other reminders to denote the affected versions. 
(3) These changes are more readily detectable by static analysis methods, as only potentially affected versions need to be scrutinized.
Above all, we systematically define the types of Incompatible APIs within the Android framework as follows.

CIs are caused by different behaviors $B$ of the same Android APIs across different versions.
Therefore, the behaviour of an API method in version $x$ can be represented as:
\begin{equation*}
\centering
    B_x = (R_x, E_x, S_x)
\end{equation*}
where $R$ stands for return (including its type and value), $E$ stands for exception handling (if any), and $S$ stands for the API signature. Therefore, CIs can be formally represented as:
\begin{equation*}
\centering
\forall x (B_x \neq B_{x+1}) \implies \exists {CI}_{(S_x, S_{x+1})} \in \mathbf{CI_{(x, x+1)}} 
\end{equation*}
Based on our empirical observation, we categorize incompatible APIs into "Signature" and "Semantic" based on the location of the vital change, i.e., whether the root reason locates in the API signature or the API implementation. 

\circled{1} Signature incompatible API means that the issue arises from changes in the API's signature, such as changes in its class name, method name, or parameters. These changes can directly break the compatibility as they often require changes in the calling code to adapt to the new signature.
Specifically, they are:
\begin{itemize} [leftmargin=*, noitemsep, topsep=0pt]
    \item \textbf{Type-1, API Addition.} A new API is introduced. This category captures scenarios where new functionality is introduced in the framework, potentially leading to CIs for applications that are not updated to leverage or accommodate these new methods.
    \begin{equation*}
    \centering
    \forall x (S_x \in \varnothing, S_{x+1} \notin \varnothing) \implies \exists CI_{(S_x, S_{x+1})}^{signature} \in \mathbf{CI_{(x, x+1)}} 
    \end{equation*}
    
    \item \textbf{Type-2, API Removal.} An old Method is deprecated.
     \begin{equation*}
    \centering
    \forall x (S_x \notin \varnothing, S_{x+1} \in \varnothing) \implies \exists CI_{(S_x, S_{x+1})}^{signature} \in \mathbf{CI_{(x, x+1)}} 
    \end{equation*}
    % \item \textbf{Type-3, API Changed.} The class name, method name, or parameters of an existing API is alternated.
    % \begin{equation*}
    % \centering
    % \forall S_x \notin \varnothing, S_{x+1} \notin \varnothing, S_x \neq S_{x+1}, \exists CI_{(S_x, S_{x+1})}^{signature} \in \mathbf{CI_{(x, x+1)}}  
    % \end{equation*}
\end{itemize}
Those incompatible API are more straightforward to identify and often result in more apparent failures or misbehavior in Android applications. 

%-----------------
\begin{figure}[t]
  \centering  
  \includegraphics[width=.9\linewidth]{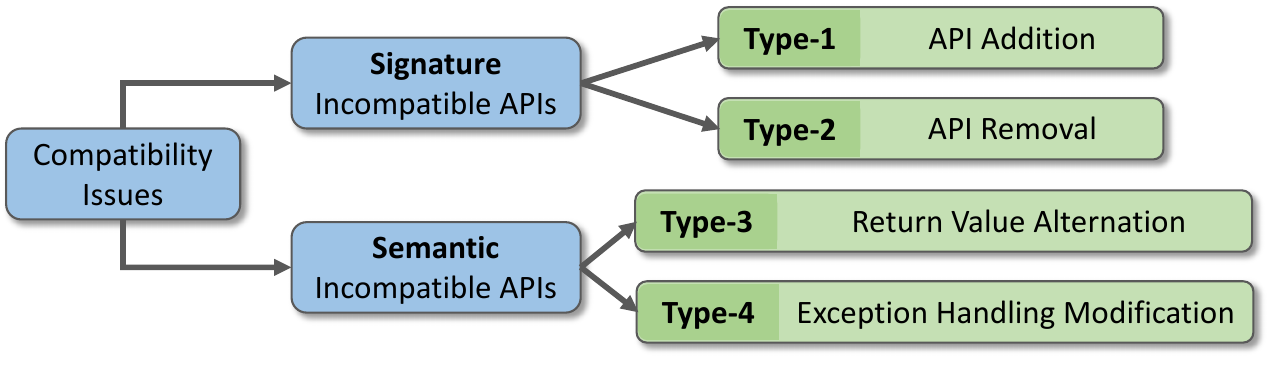}

  \caption{The taxonomy of Incompatible APIs.}
  \label{fig_CITaxonomy}
\end{figure}
%-----------------

\circled{2} Semantic incompatible API means that the issue is rooted in the change of the underlying behavior or logic of the API without altering its signature. This type of CI is more insidious as the API appears unchanged from a signature perspective, but its internal behavior have changed, leading to subtle bugs or inconsistencies in the application's behavior. Unfortunately, existing tools cannot efficiently and systematically detect CIs caused by subtle semantic changes.
In this study, we first formally define the Semantic CIs as the following two types based on API behaviours:
\begin{itemize} [leftmargin=*, noitemsep, topsep=0pt]
    \item \textbf{Type-3, Return Value Alternation.} The API potentially returns different variable types or values in the two versions.
    \begin{equation*}
    \centering
    \forall x (R_x \neq R_{x+1}, \exists B_x \neq B_{x+1}) \implies CI_{(S_x, S_{x+1})}^{semantic} \in \mathbf{CI_{(x, x+1)}} 
    \end{equation*}
    \item \textbf{Type-4, Exception Handling Modification.} The API potentially *throws different exceptions* in the two versions.
    \begin{equation*}
    \centering
    \forall x (E_x \neq E_{x+1}, \exists B_x \neq B_{x+1}) \implies CI_{(S_x, S_{x+1})}^{semantic}  \in \mathbf{CI_{(x, x+1)}}  
    \end{equation*}
\end{itemize}
Note, a CI can belong to both $CI_{signature}$ and $CI_{semantic}$.

% \circled{3} Dependency-induced incompatible API denotes that the incompatibility arises when an API's behaviour is altered due to changes in other API methods or field it depends on. 
% Such changes can lead to unexpected behaviors in applications that use the API, especially if the developers are not aware of these hidden dependencies.
% If the dependency relationship $D$ stays unchanged in version $x$ and $x+1$, then the CI can be represented as:
% \begin{align*}
% \forall D(S_x, S^\prime_x) \notin \varnothing, D(S_{x+1}, S^\prime_{x+1}) = D(S_x, S^\prime_x), B^\prime_x \neq B^\prime_{x+1}  \\
% \exists B_x \neq B_{x+1}, CI_{(S_x, S_{x+1})}^{dependency} \in \mathbf{CI_{(x, x+1)}} 
% \end{align*}

Recognizing these dependencies is crucial for developers in implementing effective CI prevention mechanisms.
Based on the definitions, we aim to obtain the full list of incompatible API signature $\mathbf{S_x}$ for $x = [4,33], x \in \mathbb{Z}$.

\subsection{LLMs in Code Analysis}
The application of pre-trained Large Language Models (LLMs) in Software Engineering (SE) tasks primarily stems from an innovative viewpoint, where a multitude of SE challenges are effectively reframed as tasks involving data, code, or text analysis~\cite{hou2023large}. This perspective has unlocked a wealth of potential breakthroughs, particularly in addressing complex and diverse SE tasks. LLMs have demonstrated significant efficacy in code-related tasks, such as summarization~\cite{sun2023automatic} (yielding an abstract natural language depiction of a code’s general functionality), and generation~\cite{yu2023codereval, liu2023your, liao2023context} (producing well-structured code and code artifacts like annotations based on users' demands). In addition to these relatively straightforward tasks, LLMs have also shown promise in more challenging areas that demand specific domain knowledge. These tasks include multi-choice code question answering~\cite{singhal2023towards, huang2021cosqa, zhao2021calibrate}, which assesses a model's ability to comprehend and analyze code snippets in a test-like format, and vulnerability detection~\cite{wang2023defecthunter, lu2024grace, senanayake2023android}, which involves identifying potential security flaws within codebases. Moreover, LLMs have been applied to bug fixing~\cite{liu2024llm, bouzenia2024repairagent, jin2023inferfix}, where they assist in identifying and correcting errors in code, showcasing their ability to not only understand but also enhance the quality and reliability of software.

Building on these breakthroughs, our study seeks to address the longstanding challenge of incompatible API detection by employing LLMs. However, we identify two primary concerns: (1) LLMs may lack the capability for fine-grained static analysis, especially without sufficient context knowledge such as individual Android API methods; (2) while LLMs cannot execute code dynamically, their ability to accurately interpret code behavior remains uncertain; and (3) when employ LLMs to conduct a binary code-related task (e.g., whether the code is vulnerable), LLMs tend to return affirmative answers, which can be mostly attributed to its hallucination. 
Despite these concerns, we investigate the potential of LLMs in this context, as detailed in Section~\ref{sec_eval_RQ1}.

% \subsubsection{CI Types - Control Flow}

% Then, from the perspective of control flow, we categorise CI into three types based on how the vital change (i.e. the root reason) affects the current method. Specifically:
% \begin{itemize} [leftmargin=*, noitemsep, topsep=0pt]
% \item \textbf{CF-1: Intra-procedure.} The change of current method leads to potential CI.
% \item \textbf{CF-2: Inter-procedure.} The change of other APIs (methods) leads to potential CI. 
% \end{itemize} 

% \xiaoyu{need a more clear definition of CI here. Step 1) Defined basic CI types based on the root causes of code evolution (i.e., signature CI --- add/delete/remove/update, semantic CI --- Return value/control depanency/exception handling).  Step 2) Defined control flow types (i.e., Intra-procedure CI --- CI appeared in the Intra-procedural statements/CI  CI appeared in the Intre-procedural statements/CI appeared in the Intra-Callback statements)} \sd{Control flow types for what?}
% \xiaoyu{Please also note that a CI may fall into several aforementioned categories.}
% \xiaoyu{move the definition of CI types from the motivation section to a "Priliminary Study Section"}
% \xiaoyu{We'd better to choose a sophisticated code example from experiments that falls into several CI categories, pointing out the detection of compatibility issues is a non-trivial task for existing state-of-the-art.}\sd{Noted.}

%-----------------
\begin{figure*}[t]
  \centering
  \includegraphics[width=0.9\linewidth]{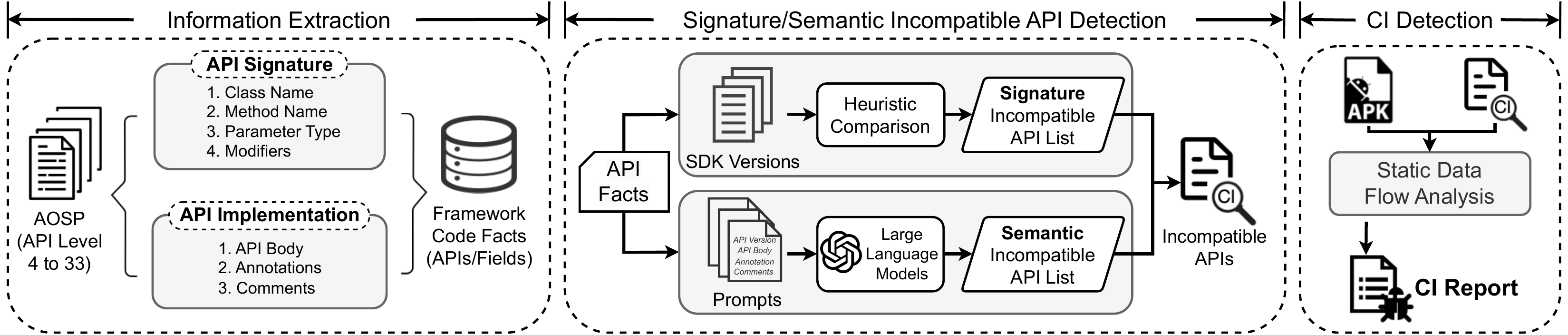}

  \caption{An overview of this study.
  }
  \label{fig_pipeline}
\end{figure*}
%-----------------

\section{Our Approach}
In this section, we introduce our novel approach to detect incompatible APIs and the consequent compatibility issues in Android frameworks. Figure~\ref{fig_pipeline} shows an overview of the working process, including the \emph{Information Extraction module}, the \emph{Incompatible API Detection} modules, and the \emph{Compatibility Issue Detection} module. 

\subsection{Information Extraction}

We first downloaded the source code of AOSP from API level 4 to 33.
All source code are obtained from \texttt{platform/frameworks/base} path.
Android APIs are mostly implemented as java methods in AOSP, so we extracted all Java methods by the steps described below.
% com.github.javaparser
% retrieve all folders get all files -> for each file, get all methods and fields -> retrieve information that we needed -> save to several csv files (each version -> 1 csv file for signature and 1 csv file for field)
Specifically, for each AOSP version, we first filtered out all non-Java files, and employed \textit{javaparser}\footnote{\url{https://github.com/javaparser/javaparser}} to retrieve the following information for all API methods:

% How to handle method overloading (Inheritance, generic type, Varargs)?
 \begin{itemize}
\item \textbf{API Signature.} API signature includes class name, method name, and parameters. It is the primary key for the following data collection. E.g., \textit{``< android.hardware.Camera.Parameters : Size getPictureSize () >''}
\item \textbf{API Body.} API body is the main functional implementation of the API, and its non-refactoring semantic change often explicitly makes the API incompatible. 
\item \textbf{API Annotation.} Method annotations in Android provide metadata about the methods, which can influence how the methods are used or interact with other components of the Android framework. E.g., the \textit{``@Deprecated''} denotes that the API is unavailable.
\item \textbf{API Comment.} API comment is another essential element when developing apps based on the Android framework. Normally, it will specifically describe the purpose of the API, its usage parameters, expected behavior, return values, and any special considerations or warnings. However, we notice that the comments were not always synchronously updated when as the API updates.
\end{itemize}

\subsection{Signature Incompatible API Detection}~\label{sec_method_signature}
% \sd{We need to higihlight that we focus on the change version instead of life span.}
API signature change is the most common reason of compatibility issues in Android applications~\cite{li2018cid, sun2022mining}.
For each Android framework version $x$, we concatenate the API method signature list and compare the list with $x+1$.
To guarantee the detection completeness, we iterated and compared every pair of neighbouring Android framework versions from version 4 to 32. 

\textbf{API Addition.} If an API method signature $S$ does not exist in Version $x$ and exists in Version $x+1$, then it is classified as “API Addition”. We obtain all API Addition by calculating the difference set of $\mathbf{S}_{x+1}$ and $\mathbf{S}_{x}$. 

\textbf{API Removal.} If an API method signature $S$ exists in Version $x$ and does not exist in Version $x+1$, then it is classified as “API Removal”. We obtain all API Addition by calculating the difference set of $\mathbf{S}_{x}$ and $\mathbf{S}_{x+1}$. 

At the end of this module, we can obtain a full list of signature incompatible APIs. Our focus of this study lies on the semantic incompatible API Detection as follows.

% \textbf{API Changed.} if a method $M$ exists in both Version $x$ and Version $x+1$ but has differences in its signature, including method full name, variables, and variable types, then it is classified as “Method Changed.” This category is essential for identifying CIs where the method's external interface or behavior has been altered between versions. Such changes can subtly affect app functionality, leading to bugs or inconsistencies that may not be immediately apparent.

\subsection{Semantic Incompatible API Detection}~\label{sec_method_semantic}
If the signature of an API stay unchanged for two continuous versions ($S_x = S_{x+1}$), then we further investigate whether the API is semantically incompatible or not. 
API body is the main functional implementation and the majority of behavioral changes-induced CIs are caused by the change of such API body.
However, API body change will not necessarily make it incompatible. A large portion of code change is simply re-factoring~\cite{liu2021identifying}, i.e its external behavior remains unchanged. The main objective of refactoring is to improve the non-functional attributes of the code, making it more understandable, maintainable, and scalable.

To better scrutinize the behaviour of APIs, we first propose a multi-label taxonomy to classify the API body change between two continuous versions as below:
\begin{itemize} [leftmargin=*, noitemsep, topsep=0pt]
    \item \textbf{Return Statement Changed.} The API potentially returns different variable types or values in the two versions. Also, a new return statement is introduced or an old return statement is deleted is also regarded as this change type.
    \item \textbf{Exception Handling Statement Changed.} The API potentially throws different exceptions in the two versions. Also, a new exception handling statement is introduced or an old statement is deleted is also regarded as this change type.
    \item \textbf{Control Dependency Changed.} The control statements, such as `if', `for', `while', or `switch', has changed; Or the statements under those control statements has changed.
    \item \textbf{Other Statement Changed.} Any statement changes that not included in return statements, exception handling statements, and statements about the control dependency.
    \item \textbf{Dependent API Changed.} The current API implementation relies on another API, and the dependent API has undergone changes, including modifications to the method name and alterations in the type or number of parameters.
\end{itemize}

Similar to the Signature Incompatible API Detection, we iterated and compared every pair of neighbouring Android framework versions from version 4 to 33, and for APIs that have same signature but different implementations, we employ pre-trained LLMs to analysis the code change and detect whether the change leads to potentially incompatible API, in which prompts are leveraged.
Prompt engineering is about carefully crafting the input text (or ``prompt'') to guide the LLMs towards producing better or more desired output. 
To achieve better performance, we maneuver the following two strategies to craft the prompt.

\noindent \textbf{Chain-of-Thought (CoT).} CoT is a technique widely used in working with LLMs like chatGPT~\cite{wei2022chain, cheng2023prompt}. It involves prompting the model to generate a sequence of explicit reasoning steps leading to an answer or conclusion. 
The code change is vital to the potential change of API's behaviour. In this task, the LLM is first asked to analysis the code change between two versions, and then detect whether the API is semantically incompatible on these two versions and which type it belongs to. 

\begin{figure}[t]
  \centering
  \includegraphics[width=.95\linewidth]{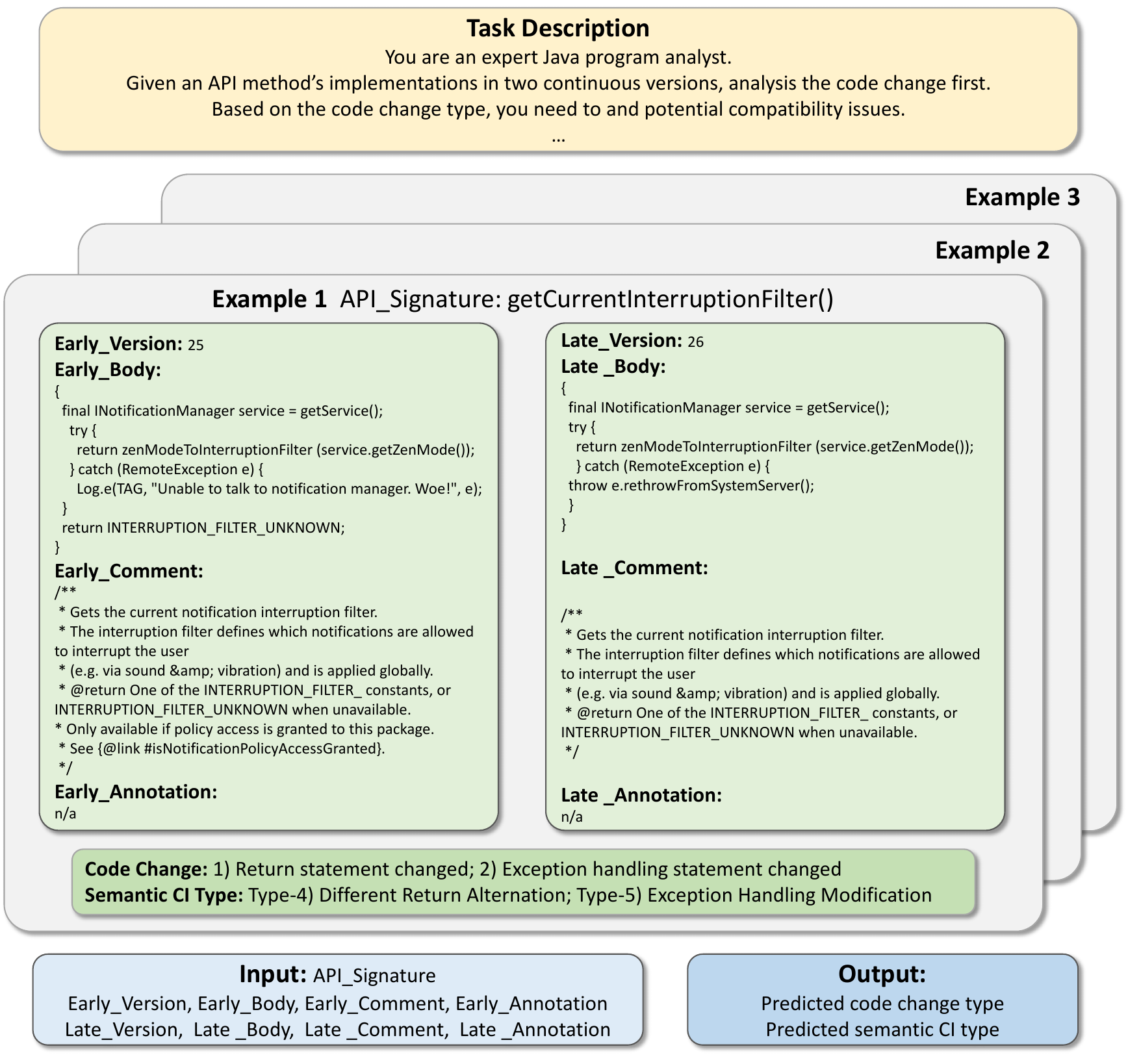}

  \caption{The design of prompt.}
  \label{fig_prompt}
\end{figure}

\noindent \textbf{In-context Learning.} Much previous work has shown that LLMs are competent on code-related tasks, and they can achieve better performance with the help of in-context learning~\cite{brown2020language, huang2022se, cheng2023prompt}.
Specifically, in addition to input a task description, some examples are more than beneficial for LLMs to complete the task.
Thus, our semantic incompatible API detection prompt consists of a task description (the light yellow box), three demonstration examples (the green boxes), and the input template (the light blue box), as shown in Figure~\ref{fig_prompt}.
Specifically, for each example, the inputs are the API’s signature, API bodies, API annotations, and API comments in an earlier version and a later version, respectively. The outputs are the code change type as an intermediate result, and the type of semantic incompatible API.

%-----------------
%-----------------
% Listing~\ref{lst_CISemantic_2} shows an example of how the semantic CI detection works. For \texttt{onSearchRequested()}, as its condition statement of \texttt{if} has changed, the left value of the condition has an additional requirement (not only not equal to~\texttt{Configuration.UI\_MODE\_TYPE\_TELEVISION} but also not equal to~\texttt{Configuration.UI\_MODE\_TYPE\_WATCH}) in the newer version, which would result in different 

Listing~\ref{lst_CISemantic_2} illustrates the concise development history of the API \emph{getNotificationPolicy()} source code. This API was incorporated into the Android framework starting from API level 23. Throughout the evolution of the Android framework, the implementation of \emph{getNotificationPolicy()} undergoes rapid modifications at API level 24, specifically in no longer returning null (line 9), indicating potential compatibility concerns due to shifted semantics.

\begin{lstlisting}[caption = {Code example of API getNotificationPolicy()}, label={lst_CISemantic_2}]
<android.app.Activity: boolean getNotificationPolicy()>
// Version 23
{
    INotificationManager service = getService();
    try {
        return service.getNotificationPolicy(mContext.getOpPackageName());
    } catch (RemoteException e) {
    }
    return null;
}
// Version 24
{
    INotificationManager service = getService();
    try {
        return service.getNotificationPolicy(mContext.getOpPackageName());
    } catch (RemoteException e) {
        throw e.rethrowFromSystemServer();
    }}
\end{lstlisting}

\subsection{Incompatibility Issue Detection}~\label{sec_method_semantic}
The ultimate goal of this work is to detect API-induced compatibility issues. To this end, we implemented the \emph{Incompatibility Issue Detection} module that extends state-of-the-art tool
CiD~\cite{li2018cid} to facilitate incompatibility detection. CiD facilitates incompatibility detection by leveraging data flow analysis to verify if an API is called under appropriate SDK protection conditions.

CiD is a state-of-the-art tool that provides a highly precise static analysis model. However, CiD only handles syntactic changes (i.e., API signature changes), overlooking semantic-induced incompatibility~\cite{sun2022mining}, which leads to false negatives. Thus, in this work, we extend CiD by supporting the declaration of semantically evolved APIs to pinpoint incompatibility originating from both syntactic and semantic changes, rather than focusing on specific sources of interest. Specifically

In the implementation of CiD, it uses the \emph{android\_api\_lifetime.txt} file to configure all signature\-incompatible APIs and pass these APIs to the data flow process to determine whether they are protected by the appropriate SDK version. To this end, we created a new configuration file, \emph{android\_api\_lifetime.txt}, which passes the identified semantically incompatible API list to the subsequent data flow analysis process. We also adapted CiD's implementation to handle semantic APIs. In this way, both syntactic and semantic incompatibilities can be detected.

% \sd{adding description here @Xiaoyu.}\xiaoyu{done}

\section{Evaluation}
Our evaluation aims to answer the following research
questions.

\begin{description}
\setlength\itemsep{0.05em}

\item[RQ1] \emph{Is the proposed approach effective on identifying semantic incompatible APIs in Android framework?} We evaluate the performance on the crafted benchmark dataset and investigate the effect of different approach settings on API code changes and semantic incompatible API detection.
%~\pei{not quite sure what do you mean. Do you mean different LLMs on different API code changes? Then why emphasize on semantic incompatible API detection? Maybe we could replace the LLMs with the first module of IntelliCiD?} 

\item[RQ2] \emph{What is the status quo of incompatible APIs in Android framework?} We employ our approach to detect all potential incompatible APIs in Android framework from version 4 to 33. Also, we conduct comprehensive statistic analysis.

\item[RQ3] \emph{What is the performance on real-world applications in detecting compatibility issues?} Given the incompatible APIs, we evaluate the capability of proposed approach in detecting compatibility issues in a large-scale real-world applications.
\end{description}

%==================
%
\begin{table*}
\caption{The performance of Semantic Compatibility API Detection module. All results of the GPT-4 are underlined. }
\label{tab_semantic_CI_result}
    \centering
    \begin{tabular}{ccl|ccc|ccc}
    \toprule
        \multicolumn{3}{c|}{\textbf{Experiment Settings}} &\multicolumn{3}{c|}{\textbf{Code Change Type}} &\multicolumn{3}{c}{\textbf{Semantic Compatibility Type}}\\
        \midrule
        No.&Foundation LLM & \multicolumn{1}{c|}{API Information} & Precision & Recall  & F1-score & Precision & Recall  & F1-score\\     
         \midrule
         1 & GPT-3.5 & body+annotation & 0.681 & 0.549 & 0.586 & 0.653 & 0.645 & 0.644 \\
         2 & GPT-4  & body+annotation & 0.773& 0.742 & 0.743 & \underline{\textbf{0.830}} & \underline{\textbf{0.837}} & \underline{\textbf{0.830}}  \\
         3 & Mistral & body+annotation & 0.728 & 0.612 & 0.641 & 0.727 & 0.721 & 0.713\\
         4 & Cohere & body+annotation & 0.720 & 0.630 & 0.643 & 0.749 & 0.741& 0.739\\ 
         \midrule
         5 & GPT-3.5 & body+annotation+comment & 0.692 & 0.561 & 0.598 & 0.602 & 0.598 & 0.592 \\
         6 & GPT-4  & body+annotation+comment & \textbf{0.784} & \textbf{0.755} & \textbf{0.747} & \underline{0.792} & \underline{0.792} & \underline{0.787} \\
         7 & Mistral & body+annotation+comment & 0.695 & 0.540 & 0.584 & 0.703 & 0.712 & 0.699\\
         8 & Cohere & body+annotation+comment & 0.703 & 0.599 & 0.619 & 0.731 & 0.721 & 0.720\\
         \midrule
         % 2 & GPT-3.5 & AST+annotation & 0.598 & 0.445 & 0.484 & 0.634 & 0.619 & 0.623 \\
         9 & GPT-3.5 & body+annotation+AST & 0.676 & 0.525 & 0.565 & 0.751 & 0.728 & 0.735 \\
         %4 & GPT-3.5 & body+annotation+comment & 0.692 & 0.561 & 0.598 & 0.602 & 0.598 & 0.592 \\
         %5 & GPT-3.5 & AST+annotation+comment & 0.473 & 0.311 & 0.352 & 0.604 & 0.590 & 0.594 \\
         10 & GPT-3.5 & body+annotation+comment+AST & 0.599 & 0.444 & 0.486 & 0.706 & 0.682 & 0.690 \\
         %7 & GPT-4  & body+annotation & 0.773& 0.742 & 0.743 & \textbf{0.830} & \textbf{0.837} & \textbf{0.830}  \\
         %8 & GPT-4  & AST+annotation & 0.709 & 0.659 & 0.661 & \textbf{0.760} & \textbf{0.746} & \textbf{0.748}  \\
         11 & GPT-4 & body+annotation+AST & 0.731 & 0.713 & 0.696 & \underline{0.777} & \underline{0.788} & \underline{0.778} \\
         %10 & GPT-4  & body+annotation+comment & \textbf{0.784} & \textbf{0.755} & \textbf{0.747} & 0.792 & 0.792 & 0.787 \\
         %11 & GPT-4  & AST+annotation+comment & 0.698 & 0.668 & 0.655 & 0.728 & 0.726 & 0.721 \\
         12 & GPT-4 & body+annotation+comment+AST & 0.726 & 0.680 & 0.676 & \underline{0.761} & \underline{0.774} & \underline{0.761}\\
         % 13 & mistral-large-2402 & body+annotation & 0.728 & 0.612 & 0.641 & 0.727 & 0.721 & 0.713\\
         % 14 & mistral-large-2402 & AST+annotation & 0.600 & 0.530 & 0.544 & 0.638 & 0.667 & 0.642\\
         % 15 & mistral-large-2402 & body+annotation+comment & 0.695 & 0.540 & 0.584 & 0.703 & 0.712 & 0.699\\
         % 16 & mistral-large-2402 & AST+annotation+comment & 0.592 & 0.486 & 0.514 & 0.658 & 0.682 & 0.660\\
         % % 17 & Jurassic-2 & body+annotation & 0.495 & 0.347 & 0.386 & 0.582 & 0.578 & 0.573\\
         % % 18 & Jurassic-2 & AST+annotation & 0.337 & 0.236 & 0.259 & 0.668 & 0.674 & 0.662\\
         % % 19 & Jurassic-2 & body+annotation+comment & 0.457 & 0.320 & 0.356 & 0.503 & 0.498 & 0.506\\
         % % 20 & Jurassic-2 & AST+annotation+comment & 0.336 & 0.249 & 0.269 & 0.643 & 0.640 & 0.633\\
         % 21 & Cohere & body+annotation & 0.720 & 0.630 & 0.643 & 0.749 & 0.741& 0.739\\
         % 22 & Cohere & AST+annotation & 0.634 & 0.509 & 0.532 & 0.715 & 0.712 & 0.707\\
         % 23 & Cohere & body+annotation+comment & 0.703 & 0.599 & 0.619 & 0.731 & 0.721 & 0.720\\
         % 24 & Cohere & AST+annotation+comment & 0.593 & 0.505 & 0.512 & 0.642 & 0.632 & 0.633\\
         \midrule
         13 & GPT-3.5 & body+annotation & \multicolumn{2}{c}{no code change type} && 0.674 & 0.659 & 0.660\\
         14 & GPT-4 & body+annotation &  \multicolumn{2}{c}{no code change type} && \underline{0.752} & \underline{0.773} & \underline{0.754}\\
    \bottomrule
    \end{tabular}
\end{table*}

% GPT-4 no comment
%Precision, Recall, F1-score
% change: 0.7728013029315962 0.742399565689468 0.7344656429346982
% CI: 0.8306188925081434 0.8371335504885994 0.8295331161780672

% GPT-4 comment
% Precision, Recall, F1-score
% change: 0.7842019543973942 0.7552660152008686 0.7466263378315494
% CI: 0.7915309446254072 0.7915309446254072 0.787187839305103

% gpt3.5 no comment
% Precision, Recall, F1-score
% change: 0.6807817589576547 0.5492399565689469 0.5857297968047157
% CI: 0.6530944625407166 0.6449511400651465 0.6438653637350704
\begin{table}
\caption{The performance of semantic compatibility API detection despite the semantic compatibility type. \textit{Comment} means whether the API comment is encoded into the prompt to LLMs, \textit{Accuracy} reflects the general semantic compatibility APIs detection capability, and \textit{Success Rate} denotes how many such APIs can be detected. All results are run by GPT-4.}
\label{tab_semantic_CI_result_overall}
    \centering
    \begin{tabular}{c|ccc}
    \toprule
        Samples & \textbf{Comment}
        &\textbf{Accuracy} & \textbf{Success Rate}\\     
         \midrule
        \multirow{2}{*}{Whole Dataset} & without & \textbf{85.3\%} & 89.1\%\\ 
        & with  & 81.1\% & \textbf{89.4\%} \\
        \midrule
        \multirow{2}{*}{Same Comment} & without & \textbf{86.1\%} & \textbf{93.1\%} \\
        &with  & 79.2\% & 89.7\%  \\
    \bottomrule
    \end{tabular}
\end{table}

% tp fp fn tn
% gpt4 comment
%126 43 19 123 
% same comment
% 54 16 4 26

% gpt4 no comment
%122 26 15 140
% same comment
% 52 7 6 35

% gpt3.5
% 68 21 66 141
% same comment
% 27 6 28 34
%
%==================

\subsection{RQ1: Employing LLMs on Semantic Incompatible API Detection}~\label{sec_eval_RQ1}
As discussed in Section~\ref{sec_motivation_problem}, semantic CI detection is a \textbf{critical} yet remains as an \textbf{unresolved} challenge in Android application development. 
A pivotal initial step in implementing prevention tactics, such as version checking, is identifying which APIs require scrutiny. 
In this section, we evaluate the performance of proposed framework, focusing on its effectiveness of leveraging LLMs for detecting semantic incompatible Android APIs.

\subsubsection{Benchmark Dataset}
Detecting semantic incompatible APIs presents a significant challenge due to their inherent complexity. Currently, there is no established benchmark dataset for this task. 
To quantitatively evaluate the capability of our proposed framework and assist the following researchers, we manually crafted a semantic CI detection benchmark dataset as described below.

First, we obtained the list of APIs whose signatures stay unchanged but other information (API bodies, comments, or annotations) changed for two continuous versions ($S_x = S_{x+1}$), from level 4 (Android 1.6) to 33 (Android 13).
Then, we randomly sampled 308\footnote{The sample size is determined based on a confidence level at 95\% and a confidence interval at 5 ~\cite{israel1992determining}} instances from the list, and manually examine whether the API is semantic incompatible on between versions.
Specifically, we recruited two annotators with at least three-year programming experience and one year Android-related analysis experience, to label the benchmark data. Specifically, they were asked to read the extracted information of the API comparison, judge the API body change type (or no change), and scrutinize the semantic incompatibility type (if any), individually. 
For any disagreement, two annotators discussed and agreed on the same answer, and if the disagreement persisted, an author (a senior researcher) joined the discussion to facilitate a resolution. 
The Krippendorff's Alpha~\cite{hayes2007answering} (\textit{a.k.a} K-Alpha) is used to reflect the inter-rater reliability for multi-label classification task. The K-Alpha is $\alpha = 0.81$ for the initial manual labelling, showing substantial level of agreement.
% 52 initial code change difference out of 208. 
% 27 initial semantic CI difference out of 208.

\begin{figure}[h]
  \begin{minipage}{0.48\linewidth}
    \centering
    \includegraphics[width=\linewidth]{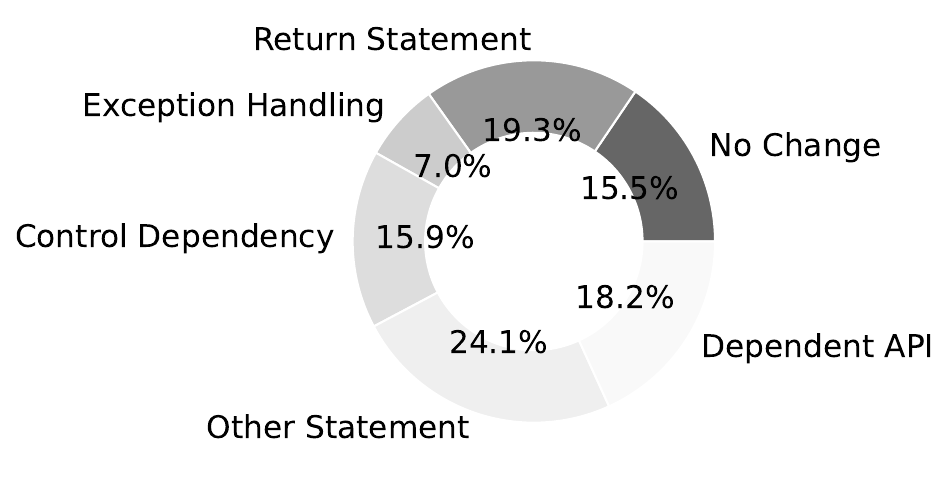}
    \caption{The distribution of code change types.}
    \label{fig_body_change_type}
  \end{minipage}%
  \hfill
  \begin{minipage}{0.48\linewidth}
    \centering
    \includegraphics[width=\linewidth]{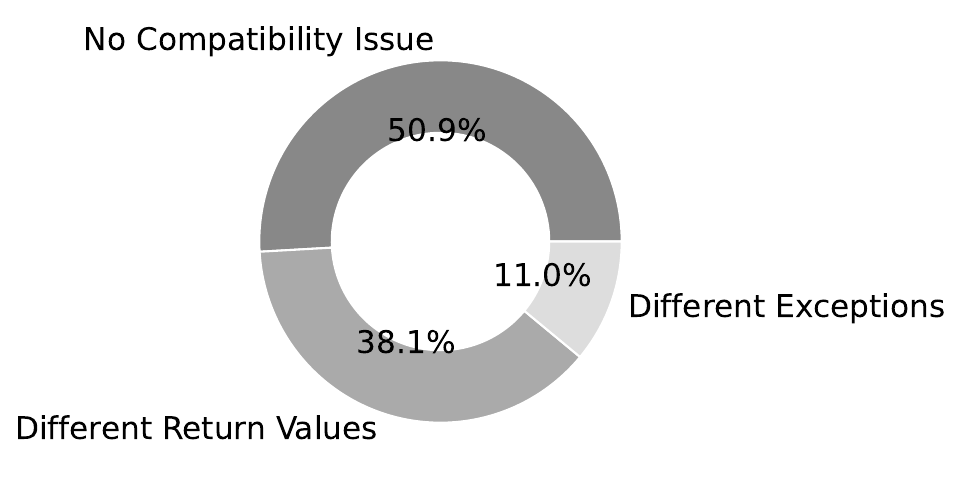}
    \caption{The distribution of compatibility issue types.}
    \label{fig_compatibility_issue_type}
  \end{minipage}
\end{figure}

%In our data annotation results, we observed that the majority of data exhibit an intra-procedure control flow type, with inter-procedure type following closely. 
Figure~\ref{fig_body_change_type} shows the distribution of code change types. We observed that "Other Statement Changed" (24.1\%) is the most common code change type compared to other classes. Such code changes occur in non-critical statements and are likely part of code re-factoring, leaving the external behaviors unchanged. Meanwhile, 15.5\% of the samples have "No Change," which means that only their API comments or annotations were updated during evolution.
Return Statement change and Control Dependency change contribute 19.3\% and 15.9\% of the total samples, respectively.

Figure~\ref{fig_compatibility_issue_type} illustrates the distribution of semantic incompatible API types. 
Notably, the dataset indicate that more than half of sampled APIs do not have potential semantic incompatibility issues (58.9\%). Additionally, 38.1\% of semantic incompatible APIs are caused by Return Value Alternation, and the rest 11.0\% APIs are caused by Exception Handling Modification. The benchmark dataset is available in our artifact package.

% \sd{The basic analysis of our benchmark dataset.}

% \begin{figure}[h]
%   \centering  
%   \includegraphics[width=.9\linewidth]{figures/APINumbers.pdf}

%   \caption{The number of provided APIs in Android frameworks over version 4 to 33.}
%   \label{fig_APINumbers}
% \end{figure}

\subsubsection{Semantic Incompatible APIs}
After obtaining the benchmark dataset, we ran the proposed approach on it. Table~\ref{tab_semantic_CI_result} presents the results and comparative analysis of the semantic compatibility detection module in the proposed framework.
In optimal settings, the framework demonstrates robust performance in classifying semantic incompatible API types, with a precision of 0.830, a recall of 0.837 and a F1 score of 0.830.
As for the code change type, the performance is lower on all metrics for this intermediate step.

\textbf{[Foundation LLM.]} 
We selected four LLMs as the foundation model of our proposed approach: GPT-3.5~\cite{openai_models}, GPT-4~\cite{openai_models}, Mistral~\cite{mistral_ai_getting_started}, and Cohere~\cite{cohere_command_r_plus}. 
The GPT series are widely considered the most powerful LLMs, and studies have shown their superiority on almost all code-related tasks~\cite{hou2023large}. Mistral is one of the cutting-edge open-weights models (outperforms Llama 34B), and we use \textit{Mistral Large} model. 
Cohere is an LLM series supports Retrieval Augmented Generation (RAG), and we use \textit{command-r-plus} with 104B parameters, which is one of the largest LLM currently.
We observed that GPT-4 achieves the best results in both Code Change type and Semantic Compatibility type classifications, followed by Cohere, Mistral, and GPT-3.5. We also noticed that the performance of Code Change type, the intermediate step, is correlated to the performance of Semantic Compatibility.

\textbf{[API Comments.]} Intuitively speaking, providing more information into AI models often leads to better performance, and it applies to the Code Change type classification. The performance of GPT-4 slightly increases on all metrics. 
Surprisingly, we discovered that providing the API comments to LLMs actually harms the performance, contrary to our expectations.
The f1-score drops from 0.830 to 0.787 for GPT-4 on the Semantic Compatibility type classification.
This counter-intuition phenomenon might stem from the LLM's interpretation of the API comments.
In real-world scenarios, Android developers often refer to API comments for a quick understanding of its functionality and behaviours. 
However, discrepancies arise when the API method body is altered but its comment remains unchanged, leading to potential misguidance to developers.
As the LLMs, mimicking human comprehension, tend to generate outputs that closely align with the provided information, it may lead to the aforementioned phenomenon.
To further verify our assumption, we took a closer inspection, and approximately one third (32.9\%) of the samples in the benchmark dataset exhibit this inconsistency. 
Table~\ref{tab_semantic_CI_result_overall} reveals that while API comments marginally impact the success rate, they impede the framework's ability to accurately determine whether the API is semantic incompatible (-4.2\%). 
This effect is more pronounced in APIs with unchanged comments but altered bodies (denoted as \textit{Same Comment}), where the inclusion of API comments reduces accuracy and success rate by 6.9\% and 3.4\%, respectively.

\textbf{[Abstract Syntax Tree (AST)]} An abstract syntax tree (AST) is a tree representation of the abstract syntactic structure of source code.
We here to explore whether providing the AST information, a symbolic explanation of Java code, of API body can help LLM understanding. 
We manually implemented a code-AST converter as the following two parts: node classes definition and method body parsing. An example of code-AST pair is listed in Listing~\ref{lst_ast}. 
\begin{lstlisting}[caption = {Code example of API getPictureSize()}, label={lst_ast}]
// Version 5, <android.hardware.Camera.Parameters: Size getPictureSize()>
{
    String pair = get(KEY_PICTURE_SIZE);
    return strToSize(pair);
}
// Version 5, <android.hardware.Camera.Parameters: Size getPictureSize()> in AST.
MethodDeclaration(method_body, [Statement({, []), AssignmentExpression(=, [VariableReference(String pair, []), Expression(get(KEY_PICTURE_SIZE), [])]), Statement(return strToSize(pair), []), Statement(}, [])])  
Statement({, [])  AssignmentExpression(=, [VariableReference(String pair, []), Expression(get(KEY_PICTURE_SIZE), [])])    VariableReference(String pair, [])    Expression(get(KEY_PICTURE_SIZE), [])  Statement(return strToSize(pair), [])  Statement(}, [])
\end{lstlisting}
% % -----------------------------------------------------------
The results show that inputting the AST into LLMs will harm the performance, the F1-score drops from 0.830 to 0.778 (-6.3\%) for GPT-4. 
LLMs are mostly trained on large corpse of natural language and normal code. 
Thus, AST in a symbolic format might might confuse the LLMs to functionally undertake the code understanding and incompatible API detection tasks.

\textbf{[Chain-of-Thought.]} The experiments shows that introducing chain-of-thought strategy can greatly enhance its capability on semantic compatibility classification, underlined by the F1-score improvement from 0.754 to 0.830 for GPT-4.

In conclusion, our framework demonstrates a commendable overall performance on semantically incompatible API classification, with an F1-score as 0.83.
As for binary semantic compatibility detection, our approach achieves an accuracy of 85.3\% and a success rate of 89.1\%. In more challenging scenarios, such as detecting semantic compatible APIs in the presence of unchanged comments and modified API bodies, our framework still shows better efficacy, evidenced by an accuracy of 86.1\% and a success rate of 93.1\%.

\find{\textbf{Answers to RQ1:} 
Our approach shows a strong overall performance with an accuracy of 85.3\% and a success rate of 89.1\% in semantic compatibility API detection, and 0.830 F1-score in semantically incompatible API classification.}

\subsection{RQ2: Statistic Analysis of Incompatible APIs}

As the proposed approach has achieved decent performance on a manual craft benchmark dataset, we then employed it to detect all potential incompatible APIs in Android framework from version 4 to 33, and the results are reported as below.

\subsubsection{Signature Incompatible APIs}
We first delve into the statistical analysis of API siganture changes occurring beyond consecutive releases. As illustrated in Figure 1, for each API level, there is a noteworthy portion of APIs has been either newly added or deleted from the prior API level. The number of our detected Signature incompatible APIs is 42,937, including addition and removal.
Utilizing such API methods may lead to compatibility issues, either due to their absence in the on-device Android stack or their removal from the public API due to insecure or non-robust behavior. Additionally, our examination of API evolution brings to light instances where certain APIs are removed from and subsequently reintroduced into the Android framework code. Such modification scenarios can give rise to latent compatibility issues, often accompanied by significant changes in API behavior. 

The prevalence of compatibility issues across diverse API levels underscores the potential hazards associated with utilizing such APIs. Inadequate verification of the SDK version by developers may give rise to exceptions within the APK, culminating in potential application crashes.

% As illustrated in Figure~\ref{fig_APINumbers}, we found that
% the number of newly added APIs of each API level vary from 
% 1,000 to 3,000, while the number of deleted APIs are ranging from 100 to 1,000. It is also worth mention that the total number of APIs experiences a more rapid growth, particularly after version 21 (Android 5.0). On average, as the version number increases, there is a greater likelihood of encountering more signature incompatible APIs.

% Thus, effective management of this situation is crucial. 

% As versions evolve, the dynamic characteristics of added and removed APIs become more apparent. 

% Developers should be adept at handling this increased variability and adjusting their strategies to adapt to the constantly evolving API environment.

\subsubsection{Semantic Incompatible APIs}
We found that among a total of 10,675 sets of APIs that have changed from API level 4 to 33, 5,481 instances of API modifications were identified as potential contributors to compatibility issues. Specifically, it was observed that 4,052 APIs were associated with compatibility issues caused by \textbf{Return Value Alteration}. Furthermore, 714 APIs were implicated in compatibility issues caused by \textbf{Exception Handling Modification}. It is also interesting to find that there are 715 APIs were concurrently identified as influencers of compatibility issues caused by both \textbf{Return Value Alteration} and \textbf{Exception Handling Modification}.

Among the remaining 5,194 APIs, with the exception of five instances deemed to cause compatibility issues under unique circumstances, the consensus was that the majority of them did not induce compatibility issues. A nuanced analysis conducted by the detection mechanism revealed that Change Type 1 (Return Statement Altered) manifested in 191 APIs, Change Type 2 (Exception Handling Statement Modified) was evident in 35 APIs, Change Type 3 (Control Dependency Altered) occurred in 1,144 APIs, Change Type 4 (Other Statement Modification) manifested in 3,474 APIs, and Change Type 5 (Dependent API Modification) was discerned in 1,833 APIs. Additionally, 345 APIs were determined to have remained unaltered between the two versions.

\begin{figure}
    \centering
    \setlength{\belowcaptionskip}{-2pt}
    \setlength{\abovecaptionskip}{-2pt}
    \includegraphics[width=1\linewidth]{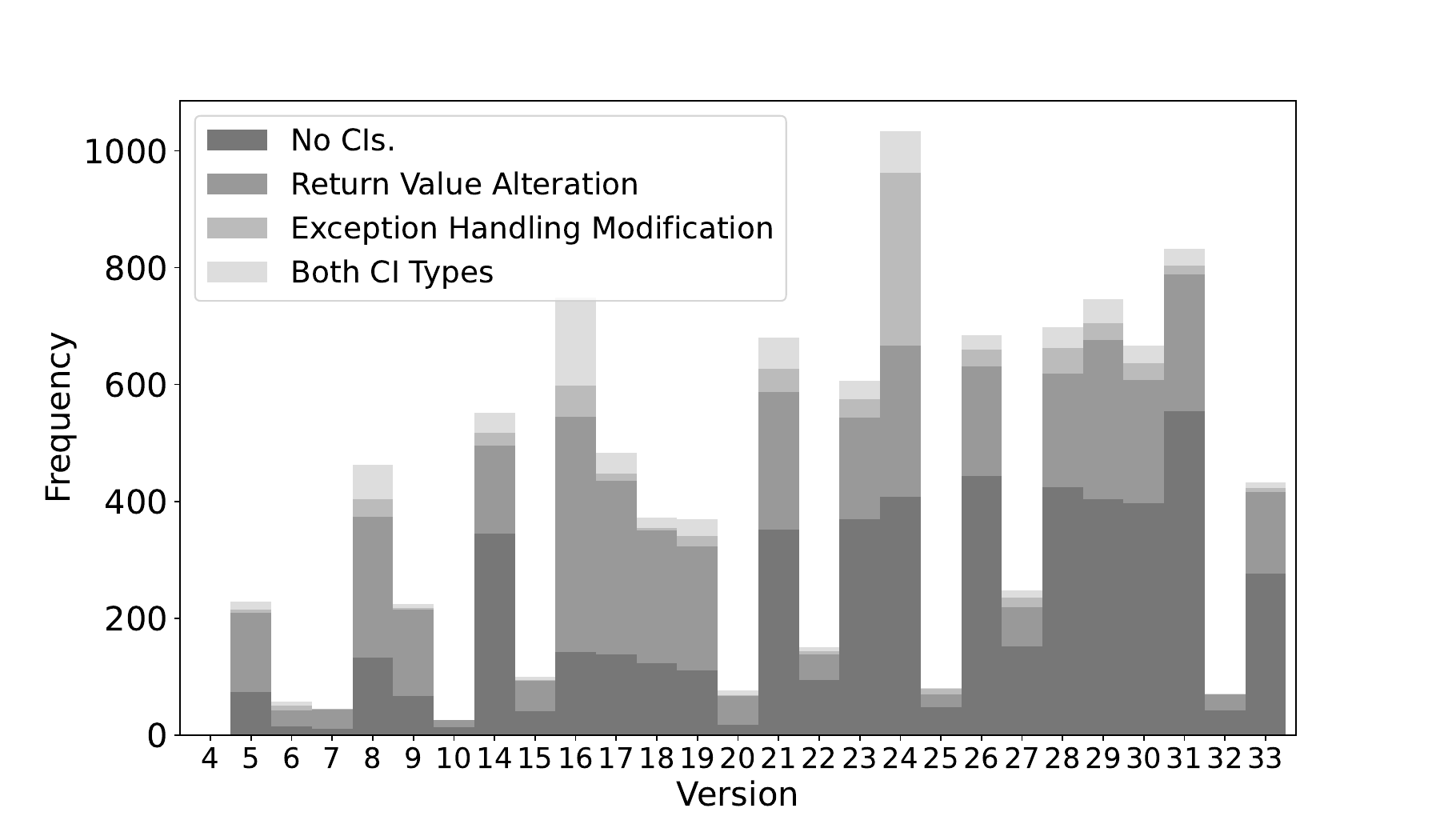}
    \caption{Detected incompatible API types generated by semantic changes between versions from 4 to 33}
    \label{fig_enter_label}
\end{figure}
Figure~\ref{fig_enter_label} further shows the number distribution of each incompatible API type detected in the semantic changes across each API level. According to the figure, in most cases, compatibility issues are caused by return statement changed. Interestingly, before version 21, the number of APIs semantic changes are relatively less than those after version 21. However, the semantic changes of APIs that will lead to compatibility issues account for a small number, and more APIs will not lead to compatibility issues even though there exist semantic changes. 
% We can also observe that the update at API level 23 to 24 is a massive change. 
The transition from API level 23 (Android 6.0, Marshmallow) to API level 24 (Android 7.0, Nougat) marked a significant evolution, as Google aimed to bolster security measures to protect user data and enhance device integrity~\cite{Android_version_history}.
More than 1,000 APIs have undergone semantic changes, and the compatibility issues caused by these changes are more likely to be caused by potential different exception handling. This change shows that compatibility issues are inevitable even if we remain vigilant in the development process, and it emphasizes that compatibility issues detection and reminder are indispensable and important in the development process of AOSP.

\subsubsection{Observations.}
% Based on the results of the \textit{Signature Incompatible API Detection} and \textit{Semantic Incompatible API Detection} modules, we obtained the full list of problematic APIs. 
% As we mentioned in Section~\ref{method_callgraph}~\pei{Commented out} there are more APIs are indirectly incompatible because the vital change locates in secondary API methods, and they can be traced by conducting control flow analysis. 

%-----------------
\begin{figure}[t]
  \centering 
  \includegraphics[width=\linewidth]{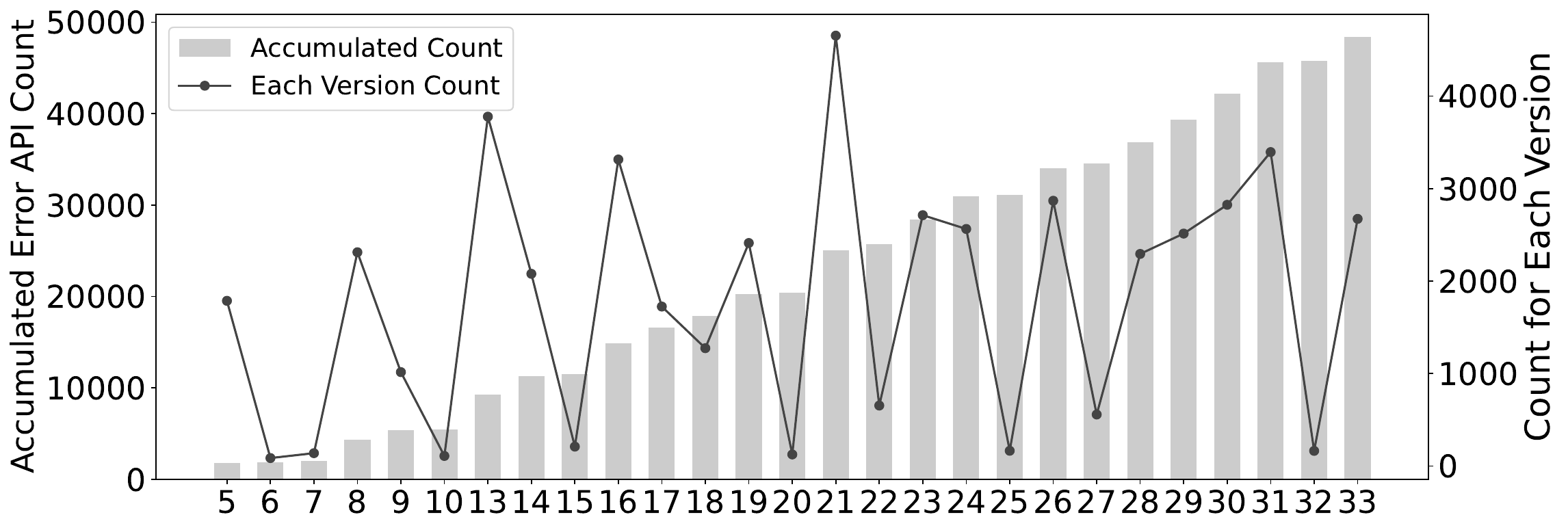}
  \caption{The Number of accumulated detected Signature and Semantic incompatible APIs ranges from version 4 to 33.}
  \label{fig_accu_APIs}
\end{figure}
%-----------------

%Our analysis revealed a large number of error APIs, specifically 35,498 (8,308 semantic + 27,190 signature) distinct error APIs and a total of 79,387 (15,494 semantic + 63,893 signature) error APIs spanning from version 4 to version 33 as detailed in Figure ~\ref{fig_Error_APIs}.
Figure~\ref{fig_accu_APIs} shows the count of detected incompatible APIs of each version (black lines) and accumulated detected incompatible APIs (grey bars).
This sizable count highlights the importance and potentials of incompatible API detection. 
An interesting observation is that when there are official version code changes in Android, such as versions 21 (version code change from KITKAT\_WATCH to LOLLIPOP)~\footnote{Google included the adoption of Material Design, enhancements in performance with the ART runtime, etc.}, and 33 (version code change from S\_V2 to TIRAMISU)~\footnote{Google provided new functionalities, enhancing security measures, etc.}, the number of error APIs often significantly increases compared to adjacent versions. This indicates that official version code changes may have a significant impact on the occurrence of error APIs.

Furthermore, it is worth noting that with the development and updates of the AOSP, the number of incompatible APIs in each version will generally increase over time. This upward trend indicates a growing number of potential incompatibility issues as the AOSP evolves. This insight is crucial for predicting and addressing compatibility challenges, highlighting the necessity of continuous vigilance and adjustment in software development practice to ensure smooth operation of applications with different AOSP versions.

% \sd{
% Thanks to the comprehensive insights from LLM, we can effectively detect these issues from APPs and provide warnings to program developers. 
% This holistic approach enables us to identify and understand the relationships between various components - call graphs, added and removed methods, and methods with semantic errors. By doing so, we have obtained a unified list of APIs, including those that may be prone to errors. This comprehensive list serves as a valuable resource for addressing and mitigating compatibility and semantic errors within the Android codebase. 

% By providing developers with a comprehensive understanding of potential issues, it enables them to take proactive measures to enhance compatibility between different software versions. This approach not only helps to solve current problems, but also contributes to a more robust and sustainable software development process. Ultimately, the goal is to facilitate smoother transitions between software versions, ensuring a seamless user experience and maintaining the integrity of Android applications over time.
% }

\find{Answers to RQ2: Based on our observation of API signature and semantic changes from API level 4 to 33, we discovered that semantic-level API alterations led to two types of compatibility issues (CIs): those arising from alterations in return values and those stemming from changes in exception handling. Additionally, our experimental results indicate that our method achieved an f1-score of 0.830 under the best experimental settings.}

\subsection{RQ3: Performance on Real-world Applications}
The objective of incompatible API analysis is to provide the necessary information for static analyzers to better identify compatibility issues in Android applications. 
In other words, armed with the incompatible API list, we resorted to supplement the state-of-the-arts in detecting a broader spectrum of compatibility issues that have been previously overlooked. To this end, we chose CiD~\cite{li2018cid} for evaluation, as it is acknowledged as the most cutting-edge static method for detecting compatibility issues in Android applications~\cite{liu2023automatically}.
However, CiD only models and compares API signatures across different SDK versions, potentially missing compatibility issues induced by API semantic differences.
To address this limitation, we enhanced CiD by incorporating the list of semantically incompatible APIs into the process of compatibility issues detection. We then conducted the comparison to show the capability of our proposed approach on real-world applications.

% \begin{figure}
%     \centering
%     \setlength{\abovecaptionskip}{0pt}
%     \setlength{\belowcaptionskip}{-4pt}
%     \subfigure[Signature CIs]{\includegraphics[width=0.15\textwidth]{figures/signature_CI_boxplot.pdf}} 
%     \subfigure[Semantic CIs]{\includegraphics[width=0.15\textwidth]{figures/allCI_boxplot.pdf}}
%     \subfigure[Sig.+Semantic CIs]{\includegraphics[width=0.15\textwidth]{figures/allCI_boxplot.pdf}} 
%     \caption{The distribution of compatibility issues per app.}
%     \label{fig:CI_per_App}
% \end{figure}

\textbf{Experimental Setup.} We randomly selected 10,000 apps, published after 2020-01-01, from AndroZoo~\cite{liu2020androzooopen} to set up the mobile app dataset of our experiment. AndroZoo is a large-scale and growing Android app collection extracted from multiple sources, including the Google Play app store.
AndroZoo has been widely employed by the Android app research community on various tasks~\cite{liu2020androzooopen, qiu2020survey, pan2023large}.
Our experiment runs on a Linux server with Intel(R) Core(TM) i9-9920X CPU @ 3.50GHz and 128GB RAM. 
The timeout setting for analyzing each app is 20 minutes. A 20-minute timeout is deemed appropriate for our study, considering that the majority of apps can undergo successful analysis within this timeframe.

\textbf{Results.} 
% Out of the 10,000 real-world applications, \tool{} successfully analyzed 8,750 of them. 
% The remaining failed cases are primarily due to exceptions encountered in Soot~\cite{vallee2010soot}, the underlying tools used by \tool{}. Thus, we reported the experimental results based on the successfully analyzed 8,750 apps. 
Overall, from the 10,000 real-world apps, we detected 280,266 compatibility issues (i.e., 144,982 signature compatibility issues and 135,284 semantic compatibility issues), \textit{w.r.t.} 6,301 distinct Android APIs. This result indicates that more than 48.27\% compatibility issues are induced by semantic change, which are non-trivial to be identified by state-of-the-arts as they are not capable of detecting compatibility issues with sophisticated implementation change. In other words, our work can supplement the state-of-the-art method in detecting a broader spectrum of compatibility issues (with an improvement of +92.3\%) that have been previously overlooked. The complete list of detected compatibility issues is available in our artifact repository.

% Figure~\ref{fig:CI_per_App} provides a summary of the distribution of the number of each compatibility issue type per app. The median number of signature, semantic and total CIs per app are xxx, xxx, and xxx respectively. 

Table~\ref{tab:CI_APIs} further summarizes the top 5 unprotected APIs that would lead to compatibility issues. 
The widespread prevalence of semantic compatibility issues suggests that our identification of an incompatible API list from the Android framework using LLM can complement existing state-of-the-art static analysis-based methods. Our proposed method can provide a more comprehensive overview of the incompatible API list, thereby enhancing the capability of these tools to detect a broader spectrum of compatibility issues.
% ~\pei{Shorten links can be generated with tools such as~\url{https://www.shorturl.at/shortener.php}}

\begin{table}[!h]
\scriptsize
\centering
\caption{Top 5 Unprotected APIs prone to compatibility issues.} 
\label{tab:CI_APIs}
\resizebox{\linewidth}{!}{
\begin{tabular}{|c |l| c | c |} 
\hline
 \textbf{CI}&  \multicolumn{1}{c|}{\textbf{API}} & \textbf{Frequency} & \textbf{Evidence} \\
 \hline
 & android.text.StaticLayout\#constructor & 1,756 & \href{https://stackoverflow.com/questions/32637224/why-does-calling-staticlayout-builder-throw-the-exception-java-lang-noclassdeffo}{Link ~\cite{StaticLayout_constructor}}\\
  & android.view.Window.Callback\#onSearchRequested & 1,624 & \href{https://github.com/greenrobot/EventBus/issues/287}{Link ~\cite{Callback_onSearchRequested}} \\
Signature & PackageInstaller.SessionInfo\#getAppPackageName & 1,613 & \href{https://stackoverflow.com/questions/26884956/how-to-install-update-remove-apk-using-packageinstaller-class-in-android-l}{Link ~\cite{SessionInfo_getAppPackageName}}\\
  % CI & android.webkit.ConsoleMessage.MessageLevel\#values  & 1,519 & 
  
  % \\
 CI & android.app.Notification.Builder\#constructor & 1,510 & \href{https://stackoverflow.com/questions/17671470/android-java-lang-noclassdeffounderror-android-app-notificationbuilder}{Link ~\cite{Builder_constructor}}\\
 & android.app.AppComponentFactory\#constructor & 1,377 & \href{https://stackoverflow.com/questions/60472222/e-loadedapk-unable-to-instantiate-appcomponentfactory-only-on-android-q-api-29}{Link ~\cite{AppComponentFactory_constructor}}\\
\hline
 & android.app.Activity\#onBackPressed & 2,402 & \href{https://github.com/organicmaps/organicmaps/issues/6692}{Link ~\cite{Activity_onBackPressed}} \\
 & android.content.res.TypedArray\#getInt & 2,166 & \href{https://stackoverflow.com/questions/31668603/java-lang-runtimeexception-cannot-make-calls-to-a-recycled-instance-type-array}{Link ~\cite{TypedArray_getInt}} \\
Semantic & android.os.PowerManager\#isScreenOn  & 1,953 & \href{https://stackoverflow.com/questions/34046862/android-powermanager-isinteractive-vs-isscreenon-bug}{Link ~\cite{PowerManager_isScreenOn}} \\
 CI & android.widget.PopupWindow\#dismiss & 1,829 & \href{https://stackoverflow.com/questions/36027819/popupwindow-dim-background-in-android-6-marshmallow}{Link ~\cite{PopupWindow_dismiss}} \\
 & android.content.res.TypedArray\#getBoolean & 1,823 & N/A\\
\hline
\end{tabular} 
}
\end{table}

% the number of resolved reflective calls is 11,646, giving a resolution rate of 89\%. This high rate experimentally shows the effectiveness
% of DroidRA in resolving reflective calls for Android apps.

We then conducted a thorough investigation to confirm if the identified incompatible APIs indeed cause problems in real-world scenarios. 
In detail, we employed the Google search engine to broadly search those compatibility issues, to check whether people have spotted and engaged discussions about them.
The search terms included the signatures of incompatible APIs (as listed in Table~\ref{tab:CI_APIs}) and keywords such as \textit{compatibility''} and \textit{android version''}. 
To ensure no relevant discussions were overlooked, we also use the statements that cause the incompatibility in the API body (e.g., the newly added exception handling statements in a later version) as the searching terms.
Then, we manually scrutinized the results from the searching.

Surprisingly, we found that 9 out of 10 incompatible APIs had associated discussions, inquiries, or issues documented on prominent tech platforms such as Stack Overflow or GitHub.
As for the only API \emph{android.content.res.TypedArray\#getBoolean}, although we did not find any relevant discussions, the implementation of this method \textbf{does} vary between API levels 20 and 21.
Specifically, as shown in listing~\ref{code:getBoolean}, in version 21, it introduces a new check for the \textit{Recycled} state and may throw the \emph{RuntimeException}, which is inconsistent with the implementation on version 20 or earlier. Hence, this API indeed poses potential compatibility issues, as it may throw an exception from version 21 while functioning without issue in versions prior to 20. In summary, the fact that we identified APIs that developers were actively complaining about in real-life situations confirms that these APIs are indeed causing significant problems. This further underscores the effectiveness of our tool.

\begin{lstlisting}[
caption={Semantic change of API \emph{TypedArray.getBoolean()}.},
label=code:getBoolean, firstnumber=1, abovecaptionskip=0pt, belowcaptionskip=0pt, aboveskip=0.5pt, belowskip=0.8pt]
// Version 20
{   index *= AssetManager.STYLE_NUM_ENTRIES;
    ...;
}
// Version 21
{if (mRecycled) {
        throw new RuntimeException("Cannot make calls to a recycled instance!");
    ...;}
    index *= AssetManager.STYLE_NUM_ENTRIES;
    ...;
}}\end{lstlisting}

\find{\textbf{Answers to RQ3:} 
Our approach generates a comprehensive list of incompatible APIs, assisting state-of-the-art static analyzers in identifying a broader range of compatibility issues, particularly those arising from significant semantic changes. Statistically, our incompatible APIs contribute to a 92.3\% improvement, previously overlooked.}

\section{Discussion}

\subsection{Implications}

\subsubsection{LLM in Code Understanding}
In the realm of LLM-based code understanding, we found that an increase in information does not necessarily lead to improved performance. 
The unchanged API comments may trick people to overlook the lurking API behaviour changes.
LLMs, much like humans, can become overwhelmed or misled by excessive or contradictory data.
Therefore, it's essential to strategically design the prompt and framework to maximize the potential of LLMs.

% \subsubsection{Capability of \tool{}} 
% \tool{} demonstrates a decent capability in detecting CIs, particularly in scenarios where the behavior of APIs changes but their comments do not. 
% This discrepancy poses a unique challenge, as developers often rely on comments for quick insights, potentially leading to misunderstandings when the API behavior has changed but the documentation has not been updated accordingly.
% Additionally, \tool{} effectively identifies more hidden situations, such as dependency-induced CIs, where the inconsistency lies not within a single procedure but in the interactions between different API methods or fields. These capabilities highlight \tool{}'s advanced analytical prowess in understanding and interpreting the complexities of API interactions and behaviors, surpassing the previous CI detection tools.

\subsubsection{CI Detection and Prevention}
Our work lays the foundation for the development of more refined static analysis tools, which can be built upon our list of identified incompatible APIs. Additionally, dynamic detection tools can utilize the types of code changes (e.g., Control Dependency Changed) and API incompatibilities we have classified (e.g., Return Value Alteration or Exception Handling Modification) to create more effective testing cases.

\subsubsection{Broader Impact} 
The broader impact of our research extends beyond just the technical realm. By improving the detection and prevention of CIs, we contribute to the overall reliability and quality of software development, particularly in the Android ecosystem.
This advancement not only aids developers in creating more robust and compatible applications but also enhances the end-user experience by reducing the likelihood of encountering software bugs or crashes. Furthermore, our approach of leveraging LLMs in code evolution analysis sets a precedent for future research, potentially leading to more innovative uses of AI techniques in addressing complex challenges within the SE field.

\subsection{Limitations}
The primary constraint of this work stems from the following two points.
First, it is widely acknowledged that static program analysis methods can encounter soundness issues. Our methodology is no different, requiring careful considerations in tackling the challenging task of extracting information from AOSP. 
% Additionally, the timeout setting inevitably rules out some deep or complex relationships from being fully analyzed, leading to potential oversights in particularly intricate API interactions. 

Second, a significant factor that limits our model’s performance is the capability of the LLMs.
Although our experiments have demonstrated decent performance on the benchmark dataset, there are inherent limitations in current LLM technologies, particularly in understanding the nuanced and context-dependent nature of code~\cite{liao2023context, hou2023large, yu2023codereval, liu2023your}, which may affect the performance of our framework in certain scenarios.

Thirdly, in our work, we employ CiD for data-flow analysis aimed at identifying compatibility issues in real-world Android applications. Nevertheless, its susceptibility to advanced programming features like reflective and native calls~\cite{samhi2022jucify} might compromise the resilience of our approach, potentially affecting its overall soundness. As part of our future endeavors, we intend to incorporate methodologies devised by fellow researchers to tackle these enduring challenges, such as integrating DroidRA~\cite{sun2021taming} to alleviate the influence of reflective calls on our static analysis approach.

% Tianchen: Some complex situations are tricky to handle. One example is the construction of method-method and method-field calling pairs, especially when parsing the child's path and type, leading to prolonged processing times. This kind of situation may result in lower efficiency, and introduce challenges in maintaining high accuracy.

\section{RELATED WORK}
% \xiaoyu{@tianchen fill up this section,  presenting existing works and what are the advantages and disadvantages compared to our approach}

% Due to the widespread fragmentation in the Android ecosystem, many device brands use different platform versions, and developers face challenges in thoroughly testing their applications to identify API compatibility related issues that may lead to application crashes and subsequently decrease the overall user experience. Recent research has explored semantic analysis and related compatibility concerns. In this section, we discuss a brief overview of significant and relevant studies.

\textbf{Semantic Code Analysis} Several current methods in software analysis primarily concentrate on program structure or external documentation. Nevertheless, it is common to overlook the semantics that are present in the source code. As noted by Kuhn et al. ~\cite{kuhn2007semantic}, in order to fully understand software, information retrieval is required to utilise linguistic information available in source code, like identifier names and comments.
%As Mahmoud et al. ~\cite{mahmoud2017semantic} pointed out, topic models are important in the field of semantic code analysis because they can extract semantic subjects from the textual content of software defects. Using models such as Latent Dirichlet Allocation (LDA), a text mining technique that extracts human-readable semantic "topics" (i.e., word clusters) from a corpus of textual documents, is one often used method. ~\cite{silva2021topic}. The work of Jipeng et al. ~\cite{qiang2022short}, which examines short-text topic modelling strategies and presents an open-source library for short-text models, is representative of the many research articles that explore the optimisation of topic models. Poursabzi-Sangdeh et al. ~\cite{Forough2021manipulating} underscored the importance of model interpretability by taking into account a variety of text mining approaches in addition to topic modelling.
Compared to traditional topic models~\cite{mahmoud2017semantic, silva2021topic, qiang2022short, Forough2021manipulating}, the more advanced tool, Large Language Models (LLMs), such as GPT-3.5 and GPT-4 ~\cite{vaswani2017attention, chen2023robust, liu2023summary} are applied to semantically analysis the code. These models have demonstrated significant potential in natural language understanding and processing programming code tasks~\cite{fang2023large}~\cite{kang2023llm}. Despite of the powerful capabilities of the LLM, the outcomes can be quite unpredictable and show non-deterministic behaviour by giving distinct codes for the same query ~\cite{ouyang2023llm}. Thus, our work aims to utilize the latest tools for semantically analyzing code, addressing compatibility issues while taking into account non-determinism at the same time.

\textbf{Android Compatibility Detecting} Various methods are used for detecting Android compatibility issues which can be roughly categorized into three types: static approaches, dynamic approaches, and machine learning approaches. 
In the realm of static approaches, Mahmud et al. ~\cite{Mahmud2021ACID} leverage API differences and conducting static analysis on the source code of Android apps to identify both API invocation compatibility issues and API callback compatibility issues. %He et al. ~\cite{he2018understanding} introduce IctApiFinder. This tool employs an inter-procedural data-flow analysis framework to effectively compute the OS versions on which an API may be invoked. Through this approach, it detects evolution-induced incompatible API usages in Android applications.
On the other hand, Wang et al. ~\cite{Wang2022Aper} adopt a dynamic approach, concentrating on identifying runtime permission misuse within Android apps. %For application developers, with the popularity of mobile applications, ensuring the correctness and reliability of their applications is the top priority. However, many applications still occasionally or frequently crash, weakening their competitive advantage. Fan et al. ~\cite{fan2018large} investigated framework-specific exceptions and enhanced a dynamic testing tool called Stoat. This effort led to the discovery of three previously unknown, confirmed, and fixed crashes in Gmail and Google+. 
Using machine learning-based classifiers, Alqahtani et al. ~\cite{Alqahtani2019MLMalware} discover malicious software on Android devices.
However, it is worth noting that each of these three approach primarily focus on syntactic changes, which are easily detectable. Unfortunately, semantic issues are often overlooked. These semantic issues which represent a deeper understanding and meaning, are not detected, making the detection less stable and reliable. Our work and contribution on semantic code analysis effectively fill this gap in current research in this field.

%Notably, there isn't much research on using machine learning approaches and LLM to detect Android compatibility issues. 
%CiD is the closest approach to ours for resolving compatibility problems caused by APIs. There are distinct differences between their approach and ours: 1) Li et al. ~\cite{li2018cid} prepares the Android data by extracting APIs and modelling their lifecycles. This method only handles additions and deletions; it ignores API changes across versions, leading to insufficient detection. In our IntelliCiD approach, the addition, deletion, and modifications of APIs (both semantic and non-semantic), as well as their corresponding comments and annotations, are taken into account, which is automatically constructed using an extraction tool and LLM semantic analysis. 2) While CiD only focuses on APIs by extracting them, our work extends its attention beyond APIs to include field changes. Modifier, type, comments and annotation modification in fields are essential in our approach for a more thorough analysis.

\section{Conclusion}

In this paper, we attempt to explore compatibility issues in the Android system by understanding the syntax and semantics involved in API evolution with the assistance of pre\-trained LLMs.
We formally define compatibility issues based on the evolution of code behavior rather than API lifespan.
We extract API syntax and semantics from the AOSP spanning API levels from API level 4 to 33. This initial step is crucial for constructing code facts (such as API signatures, API implementations, control dependencies, etc.) as they serve as the foundation for subsequent compatibility issue analysis.
We subtly employ LLMs in detecting semantic incompatible APIs, achieving 85.3\% accuracy and an 89.1\% success rate in identifying such APIs in the Android framework.
Experimental results demonstrate that our approach can enhance state-of-the-art incompatibility detection tools by identifying a larger number of compatibility issues, particularly those caused by sophisticated semantic changes.

%\textbf{Data Availability.} The source code and experimental results are all made publicly available in our artifact package: \url{https://anonymous.4open.science/r/ASE_2024_LLM_Imcompatibility_Detector-BF80}.

% \noindent \textbf{Future Work}
% % Lihong, first edition
% Our future work will mainly focus on the ability to go deep into the call chain to detect CIs, and provide code repair suggestions or the best time scheme automatically, so as to facilitate the discovery and solution of potential problems in the development process. In the future, we plan to write an LLM dedicated to API testing, which will focus on enhancing the understanding of the model, enabling real-time API change monitoring and response, broadening the coverage of its system platform, and realizing cross platform testing.

\bibliographystyle{ACM-Reference-Format}
\bibliography{8_References}

\end{document}